\documentclass[showpacs]{revtex4}
\usepackage{graphicx,amsmath,amssymb,amsfonts,bbm,latexsym,color,dcolumn,epsf}

\def\be{\begin{equation}}
\def\ee{\end{equation}}
\def\lp{\left(}
\def\rp{\right)}
\def\lb{\left[}
\def\rb{\right]}
\def\om{\omega}

\def\la{\lambda}
\def\ck{\chi_k}

\begin{document}

\title{Renormalized Stress Tensor for trans-Planckian Cosmology}
\author{D. L\'opez Nacir \footnote{dnacir@df.uba.ar}}
\author{F. D. Mazzitelli \footnote{fmazzi@df.uba.ar}}
\author{C. Simeone \footnote{csimeone@df.uba.ar}}
\affiliation{Departamento de F\'\i sica {\it Juan Jos\'e
Giambiagi}, Facultad de Ciencias Exactas y Naturales, UBA, Ciudad
Universitaria, Pabell\' on I, 1428 Buenos Aires, Argentina}

\begin{abstract}

Finite expressions for the mean value of the stress tensor
corresponding to a scalar field with a generalized dispersion
relation in a Friedman--Robertson--Walker universe are obtained
using adiabatic renormalization. Formally divergent integrals are
evaluated  by means of dimensional regularization. The
renormalization procedure is shown to be equivalent to a
redefinition of the cosmological constant and the Newton constant
in the semiclassical Einstein equations.

\end{abstract}


\maketitle

\section{Introduction}

One of the most important goals of inflationary scenarios
\cite{infl} is that they provide a causal explanation for the
large scale structure of the Universe and for the anisotropy in
the Cosmic Microwave Background (CMB). The mechanism is based in
the stretching that the exponential (or quasi exponential)
expansion produces in the physical wavelengths. Therefore, a density
fluctuation of cosmological scale today originated on scales much
smaller than the Hubble radius during inflation.

If the inflationary period lasts sufficiently long to solve the
causality and other related problems, the scales of interest today
are not only within the horizon but are also {\it sub-Planckian}
at the beginning of inflation \cite{bran1}. This fact, known as
the {\it trans-Planckian problem}, is a potentially interesting
window to observe consequences of the Planck scale physics. The
inflationary models may turn the Universe into a Planck-scale
''microscope''. For this reason, since the formulation of this
problem, many authors \cite{varios-tp} have studied the
possibility of observing signatures of Planckian physics in the
power spectrum of the CMB and in the evolution of the Universe. In
the absence of a full quantum theory of gravity, the analysis is
necessarily phenomenological. For instance, modified dispersion
relations for the modes of quantum fields might arise in loop
quantum gravity \cite{loop} or due to the interaction with
gravitons \cite{grav}. It is then important to test the robustness
of inflationary predictions under such modifications. Other
possibility is to consider an effective field theory approach in
which the trans-Planckian physics is encoded in the state of the
quantum fields when they leave sub-Planckian scales \cite{eff,star}.
One could also consider space-space or space-time
non-commutativities \cite{noncom}.

In this paper we will consider the first approach, i.e. we will
analyze quantum fields with non-standard dispersion relations in
Friedman-Robertson-Walker (FRW) backgrounds. Within this
framework, in simple models with a single quantum scalar field
$\phi$, the information on the power spectrum of the CMB is
contained in the vacuum expectation value $\langle \phi^2\rangle$. Moreover,
the backreaction of the scalar field  is contained
in the expectation value  $\langle T_{\mu\nu}
\rangle$. Note that both $\langle\phi^2\rangle$ and
$\langle T_{\mu\nu}\rangle$ are in general divergent quantities.

There is a debate in the literature about whether the backreaction
of trans-Planckian modes affects significantly the background
spacetime metric or not  \cite{tana,star,star2,kalo}. If
sub-Hubble but super-Planck modes are excited during inflation,
its energy density may be of the same order of magnitude that the
background energy density, and prevent inflation. This fact would
put a bound on the occupation numbers of the excited modes, and
therefore on the effect that trans-Planckian physics may have on
the power spectrum of the CMB. This argument has been disputed in
Ref. \cite{branma}, where the authors point out some subtleties
regarding the choice of the ultraviolet cutoff and the equation of
state of the transplanckian modes.

A consistent solution of this controversy should be based in
a careful evaluation of the expectation value of the pressure and
the energy density, and in the analysis of the solutions of the Semiclassical Einstein Equations (SEE), in which $\langle T_{\mu\nu}\rangle $ acts as a source. Any physically meaningful prediction
requires the obtention of finite quantities starting from the formal
divergent expression for $\langle T_{\mu\nu}\rangle$. In previous works \cite{branma,lemoine}, a
particular renormalization prescription has been used, which
consists essentially in subtracting the ground state energy of
each Fourier mode. This prescription may lead to inconsistencies
for quantum fields in curved spaces \cite{wald}.  The purpose of
the present paper is to carefully study this problem and provide a
correct definition of such finite quantities.

The renormalization procedure for quantum fields satisfying the
standard dispersion relations in curved backgrounds is of course
well established \cite{wald,birrell,fulling}. For example, in the
point-splitting regularization technique \cite{christ},
$\langle\phi^2\rangle$ and $\langle T_{\mu\nu}\rangle$ can be
computed in terms of the coincidence limit of the two-point
function $G^{(1)}(x,x')=\langle\{\phi(x),\phi(x')\}\rangle$ and
its derivatives. The renormalized values are obtained by using the
subtracted function $G^{(1)}_{\mathrm{sub}}(x,x')= G^{(1)}(x,x')-
G^{(1)}_{\mathrm{Had}}(x,x')$ where $G^{(1)}_{\mathrm{Had}}(x,x')$
is a two-point function with the Hadamard singularity structure
\cite{wald}, truncated at the fourth adiabatic order \cite{foot}.
The limit $x' \to x$ is taken at the end of the calculation.
Alternatively, using dimensional regularization one can work with
$x'=x$ from the beginning. This renormalization procedure is
covariant, and the divergences of $\langle T_{\mu\nu}\rangle$ can
be absorbed into redefinitions of the coupling constants of the
theory in the SEE. In order to absorb all divergences it is
necessary to include terms quadratic in the curvature in the
classical gravitational action.

The method of renormalization described above can be applied in
principle in any spacetime metric. However, in the particular case
of FRW metrics, the adiabatic subtraction is simpler and more
adequate for numerical calculations \cite{birrell,fpb74,bunch}.
Instead of subtracting the two point function, the idea is to
subtract the adiabatic expansion of the modes of the quantum
fields. Adiabatic subtraction must be complemented with a
covariant regularization, as for instance dimensional
regularization. It has been shown that this method is equivalent
to the previous one \cite{equiv}.

For a quantum field with generalized dispersion relations, the
covariance is lost unless one introduces an additional dynamical
degree of freedom $u^{\mu}$ that defines a preferred rest frame
\cite{matt}. One usually works within this preferred frame, in
which the spacetime metric has the FRW form and the additional
degree of freedom does not contribute to the energy momentum
tensor. In this particular frame, because the stress tensor of the
quantum field is the source in the SEE, we must demand that it
fulfills the same conservation equation of the Einstein tensor,
i.e. $G^{\mu\nu}_{\ ;\nu}=0$ implies $\langle
T^{\mu\nu}\rangle_{;\nu}=0$. We stress that the conservation
equation for $\langle T_{\mu\nu} \rangle$ is not necessarily valid
in other frames, since the complete energy momentum tensor may
contain an additional part coming from $u^{\mu}$.

Therefore, the renormalization should be compatible with the
structure of the SEE in the preferred frame. The divergent
contributions to be subtracted must have the form of geometric
conserved tensors, in order to be absorbed into redefinitions of
the bare constants. To ensure this, we shall follow the adiabatic
renormalization procedure described above, that is, we shall
evaluate the divergent contributions of the adiabatic expansion of
the stress tensor, and define the renormalized stress tensor as
$\langle T^{\mu\nu}\rangle-\langle T^{\mu\nu}\rangle_{\mathrm
{Ad}}$. We shall show that because fourth or higher adiabatic
order contributions are already finite for the dispersion
relations considered, no additional terms must be included in the
SEE, and only a redefinition of the cosmological constant and the
Newton constant is required.

The paper is organized as follows. In Section II we generalize the
WKB expansion to fields with a non standard dispersion relation in
an arbitrary number of dimensions. In Section III we discuss the
adiabatic renormalization of the energy momentum tensor for a
generic dispersion relation, and then compute the explicit
expressions for the dimensionally regularized counterterms in some
particular cases. We include our conclusions in Section IV. In the
Appendix we describe the simpler problem of the renormalization of
$\langle\phi^2\rangle$.

Throughout the paper we set $c=1$ and adopt the sign convention
denoted (+++) by Misner, Thorne, and Wheeler \cite{MTW}.

\section{The WKB expansion}

We  begin  by computing  the WKB expansion for the modes of  a
scalar field $\phi$ with a non standard dispersion relation. The
action of the theory is given by \cite{lemoine}: \be
S=\int d^n x \sqrt{-g}
(\mathcal{L}_{\phi}+\mathcal{L}_{cor}+\mathcal{L}_{u}),\ee where
$n$ is the space-time dimension, $\mathcal{L}_{\phi}$ is the
standard Lagrangian of a free
 scalar field
\be \mathcal{L}_{\phi}=-\frac{1}{2}\lb g^{\mu
\nu}\partial_{\mu}\phi\partial_{\nu}\phi+(m^2+\xi R)\phi^2\rb,\ee
$\mathcal{L}_{cor}$ is the corrective lagrangian that gives rise
to a generalized dispersion relation \be
\mathcal{L}_{cor}=-\sum_{s,p\leq n} b_{sp}
(\mathcal{D}^{2s}\phi)(\mathcal{D}^{2p}\phi),\ee with
$\mathcal{D}^{2}\phi\equiv\perp_{\mu}^{\lambda}\nabla_{\lambda}\perp_{\gamma}^{\mu}\nabla^{\gamma}\phi$
($\perp_{\mu\nu}\equiv g_{\mu\nu}+ u_{\mu} u_{\nu}$, where
$\nabla_{\mu}$ is the corresponding covariant derivative), and
$\mathcal{L}_{u}$ describes the dynamics of the additional degree
of freedom $u^{\mu}$ whose explicit expression is not necessary for
our present purposes.

We work with a general spatially flat FRW metric given by \be
ds^2=g_{\mu\nu}dx^{\mu}dx^{\nu}\equiv
-(u_{\mu}dx^{\mu})^2+\perp_{\mu\nu}dx^{\mu}dx^{\nu}=C(\eta)[-d\eta^2+\delta_{i
j}dx^i dx^j]\ee where $C^{1/2}(\eta)$ is the scale factor given as
a function of the conformal time $\eta$, and $u_{\mu}\equiv
C^{1/2}(\eta)\delta^{\eta}_{\mu}$.

The generalized dispersion relation takes the form \be \om^2_k=
k^2+C(\eta)\lb
m^2+2\sum_{s,p}(-1)^{s+p}\,b_{sp}\,\lp\frac{k}{C^{1/2}(\eta)}\rp^{2(s+p)}\rb,
\label{dis} \ee where  $b_{sp}$ are arbitrary coefficients, with
$p\leq s$.

The Fourier modes $\chi_k$ corresponding to the scaled field
$\chi=C^{(n-2)/4}\phi$ satisfy \be
\chi_k''+\lb(\xi-\xi_n)RC+\om_k^2\rb\chi_k=0, \label{PXXP}\ee with
the usual normalization condition \be \ck
{\ck'}^*-\ck'\ck^*=i.\label{nor} \ee Here primes stand for
derivatives with respect to the conformal time $\eta$, $R$ is the
Ricci scalar, and in the conformal coupling case we have
$\xi=\xi_n\equiv(n-2)/(4n-4)$, while $\xi=0$ corresponds to
minimal coupling. The normalization condition implies that the
field modes $\chi_k$ can be expressed in the well known form \be
\ck= \frac{1}{\sqrt{ 2 W_k}}\exp\lp -i\int^\eta
W_k(\tilde\eta)d\tilde\eta\rp .\label{chi} \ee Substitution of Eq.
(\ref{chi}) into Eq. (\ref{PXXP}) yields
\begin{eqnarray}
W_k^2 & = &
\Omega_k^2-\frac{1}{2}\lp\frac{W_k''}{W_k}-\frac{3}{2}\frac{{W'_k}^2}
{W_k^2}\rp,\label{Weq} \\
\Omega_k^2 & = & \om_k^2+\lp\xi-\xi_n\rp
CR.\label{W}
\end{eqnarray}
The non-linear differential equation for $W_k$ can be solved
iteratively by assuming that it is a slowly varying function of
$\eta$. This is the adiabatic or WKB approximation, and the number
of time derivatives of a given term is called the adiabatic order.
Thus, if we work up to the second adiabatic order (which is the
highest order which will be required; see below), we can replace
$W_k$ by $\om_k$ on the right-hand side  of Eq. (\ref{Weq}). Then,
with the use of (\ref{dis}) and (\ref{Weq}), we straightforwardly
obtain the second order solution for a generic evolution of the
scale factor:
\begin{eqnarray}
W_k^2 & = & \om_k^2+\lp \xi-\xi_n\rp(n-1)\lp\frac{C''}{C}+\frac{(n-6)}{4}\frac{{C'}^2}{C^2}\rp \nonumber\\
 &-& \   \frac{1}{4}\frac{C''}{C}\lp 1-\frac{k^2}{\om_k^2}\frac{d\om_k^2}{d k^2}\rp-\frac{1}{4}\frac{{C'}^2}{C^2}\frac{k^4}{\om_k^2}\frac{d^2\om_k^2}{{d(k^2)}^2}\nonumber\\
& +&  \frac{5}{16}\frac{{C'}^2}{
C^2}\lp1-\frac{k^2}{\om_k^2}\frac{d\om_k^2}{d k^2}
\rp^2\label{W2}.
\end{eqnarray}
where we have used that $\om_k^2/C$ is a function of $k^2/C$ to
write the temporal derivatives of $\om_k$ in terms of derivatives
with respect to $k^2$. We see that $W_k^2=\om_k^2+\epsilon_k$,
where $\epsilon_k$ is already of second adiabatic order (that is,
it includes second derivatives or the square of first derivatives
of the scale factor). In the calculations below we shall need the
squared modes $|\ck|^2=\ck\ck^*=(2W_k)^{-1}$ and, up to second
adiabatic order, we can simply work with the approximation
$(W_k)^{\pm 1}\approx (\om_k)^{\pm 1}\lb 1\pm\epsilon_k/(2\om_k^2)
\rb$.

In what follows it will be relevant to know the dependence with
$k$ of the different adiabatic orders. From Eq. (\ref{W2}) it is
clear that while the zeroth adiabatic order scales as $\om_k^2$,
the second adiabatic order scales as $\om_k^0$. Using an inductive
argument it can be shown that the $2j-$adiabatic order scales as
$\om_k^{2-2j}$.

\section{Renormalization of the stress tensor}

Motivated by the work in Ref. \cite{lemoine}, we start from the
following expressions for the vacuum expectation values of the
energy density $\rho$ and pressure $p$, which we have generalized
to arbitrary dimension $n$ and coupling $\xi$ to the scalar
curvature:
\begin{eqnarray}
\nonumber\langle\rho\rangle &=& \frac{1}{ \sqrt{C}}\int
\frac{d^{n-1}k\,\mu^{4-n}}{(2\pi\sqrt{C})^{(n-1)}} \left\{
\frac{C^{(n-2)/2}}{2}\left|\lp \frac{\chi_k}{C^{(n-2)/4}}\rp'
\right|^2 +\frac{\om_k^2}{2}\,|\chi_k|^2+\xi
G_{\eta\eta}|\chi_k|^2\right.
\\ &+&\left.\xi\frac{(n-1)}{2}
\lb\frac{C'}{C}(\chi_k'\chi_k^*+\chi_k{\chi_k'}^{*})-\frac{{C'}^2}{C^2}\frac{(n-2)}{2}|\chi_k|^2
\rb \right\},\label{RHOO}\\
\nonumber\langle p\rangle &=& \frac{1}{ \sqrt{C}}\int
\frac{d^{n-1}k\,\mu^{4-n}}{(2\pi\sqrt{C})^{(n-1)}}
\left\{\lp\frac{1}{2}-2\xi\rp C^{(n-2)/2}\left|\lp
\frac{\chi_k}{C^{(n-2)/4}}\rp' \right|^2+\xi
G_{11}|\chi_k|^2\right.\\\nonumber
 &+&\lb \lp\frac{k^2}{n-1}\rp\frac{d\om_k^2}{
dk^2}-\frac{\om_k^2}{2}\rb|\chi_k|^2-\xi(\chi_k''\chi_k^*+\chi_k{\chi_k''}^{*})+\xi\frac{
C'}{2C}(\chi_k'\chi_k^*+\chi_k{\chi_k'}^{*})\\
&-&\left.\xi\frac{(n-2)}{2}\lp\frac{C''}{C}-\frac{(8-n)}{4}\frac{{C'}^2}{C^2}\rp|\chi_k|^2\right\}.\label{PPP}
\end{eqnarray}
Here $\mu$ is an arbitrary parameter with mass dimension introduced to ensure that
$\chi$ has the correct dimensionality, and $G_{\eta\eta}$ and $G_{11}$($=G_{22}=G_{33}$) are
the nontrivial components of the Einstein tensor
\begin{eqnarray} G_{\eta\eta}&=&\ \frac{(n-1)}{4}\frac{(n-2)}{2}\frac{{
C'}^2}{ C^2}, \\
G_{11}&=&G_{22}=G_{33}=\frac{(n-2)}{2}\lb\frac{{
C'}^2}{ C^2}  \lp \frac{(n-1)}{4}-\frac{(n-4)}{2}\rp-\frac{C''}{C}\rb.
\end{eqnarray}

After introducing the form of the Fourier modes given in Eq.
(\ref{chi}) in the expressions for the vacuum expectation values
$\langle\rho\rangle$ and $\langle p\rangle$ of Eqs. (\ref{RHOO})
and (\ref{PPP}) we find
\begin{eqnarray}
\nonumber\langle\rho\rangle &=&  \frac{\Omega_{n-1}}{
2\sqrt{C}}\int \frac{dk\,k^{n-2}\,\mu^{4-n}}{(2\pi\sqrt{C})^{n-1}}\left\{\frac{[(W_{k}^2)']^2}{32 W_{k}^5}+\frac{W_{k}}{ 2}+\frac{\om^2}{ 2W_{k}}+\frac{(n-2)}{2}\lb\frac{ {C'}^2(n-2)}{16 W_{k}C^2}+\frac{ {C'}(W_{k}^2 )'}{8CW_{k}^3}\rb\right.\\
&+&\left.\xi
\frac{G_{\eta\eta}}{W_{k}}-\xi\frac{(n-1)}{2}\lb\frac{{C'}^2}{
C^2}\frac{(n-2)}{2 W_{k}}+\frac{C'}{C}\frac{(W_{k}^2)'}{2 W_{k}^3}\rb\right\},
\label{rho}
\end{eqnarray}
\begin{eqnarray}
\langle p\rangle &=&  \frac{\Omega_{n-1}}{
2\sqrt{C}}\int \frac{dk\,k^{n-2}\,\mu^{4-n}}{(2\pi\sqrt{C})^{n-1}}\left\{ \frac{[(W_{k}^2)']^2}{ 32 W_{k}^5}+\frac{W_{k}}{2}-\frac{\om^2}{2W_{k}}+\frac{(n-2)}{2}\lb\frac{ {C'}^2(n-2)}{ 16 W_{k} C^2}+\frac{ {C'}(W_{k}^2 )'}{ 8CW_{k}^3}\rb \right.\label{pe}\nonumber\\
&+&\frac{k^2}{(n-1) W_{k}}\frac{d\om^2}{
dk^2}+\xi\frac{G_{11}}{W_{k}}+\xi\lb\frac{(W_{k}^2)''}{2
W_{k}^3}-\frac{3}{4}\frac{[(W_{k}^2)']^2}{W_{k}^5}-\frac{(n-1)}{4}\frac{C'(W_{k}^2)'}{C
W_{k}^3} \rb\nonumber\\
&+&\left.\frac{(n-2)}{2}\frac{\xi}{
W_{k}}\lb\frac{C''}{C}-\frac{3}{2}\frac{{C'}^2}{C^2}\rb\right\},
\end{eqnarray}
where we have defined  the factor $\Omega_{n-1}\equiv
2\pi^{(n-1)/2}/\Gamma[(n-1)/2]$ coming from the angular
integration.

The dependence with $k$ of the $2j-$adiabatic order has been
described at the end of the previous section. From that result it
is possible to check that, for $\om_k^2\sim k^r$ with $r\geq 8$,
all contributions of second or higher adiabatic order are finite.
The divergences come only from the zeroth adiabatic terms
contained in $\langle\rho\rangle$ and $\langle p\rangle$. Instead,
in the cases $\om_k^2\sim k^6$ and $\om_k^2\sim k^4$, though no
fourth order divergencies appear, second adiabatic order terms
include, in principle, divergent contributions. Therefore we only
need to work up to second adiabatic order. Since
$W_k^2=\om_k^2+\epsilon_k$, where $\epsilon_k$ is of  adiabatic
order two, we can substitute $(W_{k}^2)'$ by $(\om^2)'$ in
equations above. Then, with the help of Eq. (\ref{W2}) and the
explicit form of $\om_k^2$ (given in Eq.(\ref{dis})), we obtain
the following expressions for the zeroth and second adiabatic
order contributions:
\medskip

\begin{eqnarray}
\langle\rho\rangle^{(0)} &=& \frac{1}{ 2\sqrt{C}}\frac{\Omega_{n-1}\,\mu^{4-n}}{(2\pi\sqrt{C})^{n-1}}\int dk\,k^{n-2} \om_k ,\,\,\,\,\,\,\,\,\,\,\,\,\,\,\,\,\,\,\,\,\,\,\,\,\,\,\,\,\,\,\,\,\,\,\,\,\,\,\,\,\,\,\,\,\,\,\,\,\,\,\,\,\,\,\,\,\,\,\,\,\,\,\,\,\,\,\,\,\,\,\,\,\,\,\,\,\,\,\,\,\,\,\label{rhoderiv}\\
\langle p\rangle^{(0)} &=& \frac{1}{
2\sqrt{C}}\frac{\Omega_{n-1}\,\mu^{4-n}}{(2\pi\sqrt{C})^{n-1}}\int
dk\frac{k^{n-2}}{(n-1)} \frac{k^2}{\om_k}\frac{d\om^2_k}{
dk^2},\label{p6}
\end{eqnarray}
\begin{eqnarray}
\langle\rho\rangle^{(2)}_{\xi=0} & = &\frac{1}{ 2\sqrt{C}}\frac{\Omega_{n-1}\,\mu^{4-n}}{(2\pi\sqrt{C})^{n-1}}\lp\frac{C'}{C}\rp^2\int dk\,k^{n-2}\left\{\frac{1}{32 \om_k}\lp 1-\frac{k^2}{\om_k^2}\frac{d\om_k^2}{d k^2}\rp^2\right.\nonumber\\
& +& \left.\frac{(n-2)}{32 \om_k}\lp n-\frac{2 k^2}{\om_k^2}\frac{d\om_k^2}{d k^2}\rp\right\},\label{rhominimcoup}\\
\langle p\rangle^{(2)}_{\xi=0}
& =& \langle\rho\rangle^{(2)}_{\xi=0} -\frac{1}{ 2\sqrt{C}}\frac{\Omega_{n-1}\,\mu^{4-n}}{(2\pi\sqrt{C})^{n-1}}\int dk\,k^{n-2}\frac{1}{2\om_k}\lp 1 - \frac{k^2}{(n-1)\om_k^2}\frac{d\om_k^2}{d k^2}  \rp \nonumber\\
&\times& \left\{ \frac{(n-2)}{4}\lp\frac{C''}{C}+\frac{(n-6)}{4}\frac{{C'}^2}{C^2}\rp+\frac{C''}{4 C}\lp 1-\frac{k^2}{\om_k^2} \frac{d\om_k^2}{d k^2}\rp  \right.\nonumber\\
&+&\left.\frac{{C'}^2}{4 C^2}\lb\frac{k^4}{\om_k^2}\frac{d^2\om_k^2}{{d (k^2)}^2}- \frac{5}{4}\lp  1-\frac{k^2}{\om_k^2}\frac{d\om_k^2}{d k^2}\rp^2\rb\right\},
\end{eqnarray}
\begin{eqnarray}
\langle\rho\rangle^{(2)}_{\xi} & = & \frac{\xi}{ 2\sqrt{C}}\frac{\Omega_{n-1}\,\mu^{4-n}}{(2\pi\sqrt{C})^{n-1}}\int dk\,k^{n-2}\left\{\frac{G_{\eta\eta}}{\om_k}-\frac{{C'}^2(n-1)}{4 C^2 \om_k}\lp n-1- \frac{ k^2}{\om_k^2}\frac{d\om_k^2}{d k^2}\rp\right\} ,\\
\langle p\rangle^{(2)}_{\xi} & =& \frac{\xi}{
2\sqrt{C}}\frac{\Omega_{n-1}\,\mu^{4-n}}{(2\pi\sqrt{C})^{n-1}}\int
dk\,k^{n-2}\left\{\frac{G_{1 1}}{\om_k}+\frac{1}{2\om_k}\lb\frac{2
C''}{C}+\frac{(n-6)}{4}\frac{{C'}^2}{C^2} \rb
\right. \nonumber\\
&\times&\lp 1- \frac{ k^2}{\om_k^2}\frac{d\om_k^2}{d k^2}\rp
+3\frac{(n-2)}{2}+\frac{3}{2}\lp  1-\frac{k^2}{\om_k^2}\frac{d\om_k^2}{d k^2}\rp^2  \nonumber\\
&+&   \left. \frac{1}{2 \om_k} \frac{{C'}^2}{C^2}\lb
\frac{k^4}{\om_k^2}\frac{d^2\om_k^2}{{d (k^2)}^2}-\frac{(n-1)}{2}
\lp 1-\frac{k^2}{\om_k^2} \frac{d\om_k^2}{d k^2}  \rp\rb
\right\}\label{pderiv},
\end{eqnarray}
which include divergences coming from different powers of the wave
vector $k$. Here the superscripts stand for the adiabatic order,
and we have separated the second adiabatic order which appears in
the minimal coupling case $\langle .\rangle_{\xi=0}$ from the one
proportional to the coupling constant $\xi$.

\subsection{Zeroth adiabatic order}
The divergences in the components of the stress tensor that come
from the zeroth order in the adiabatic expansion can be removed by
renormalization of the cosmological constant in the SEE. This can
already be verified as follows: Up to zeroth order we have that
$\langle p\rangle$ is given by Eq. (\ref{p6}), which can be recast
as \be \langle p\rangle^{(0)} =\frac{1}{
2\sqrt{C}}\frac{\Omega_{n-1}}{(2\pi\sqrt{C})^{n-1}}\int
dk\frac{k^{n-1}\,\mu^{4-n}}{(n-1)} \frac{d\om_k}{ dk}, \ee so that
one can integrate by parts to obtain \be \langle p\rangle^{(0)} =
 \frac{1}{
2\sqrt{C}}\frac{\Omega_{n-1}\,\mu^{4-n}}{(2\pi\sqrt{C})^{n-1}}\left\{
\int dk\, \frac{d}{dk}\lp\frac{\om_k\,k^{n-1}}{n-1}\rp-\int
dk\,k^{n-2} \om_k\right\}.\ee  Then, since in dimensional
regularization the integral of a total derivative vanishes
\cite{Collins},  we find that \be \langle p\rangle^{(0)} = -
\frac{1}{
2\sqrt{C}}\frac{\Omega_{n-1}\,\mu^{4-n}}{(2\pi\sqrt{C})^{n-1}}\int
dk\,k^{n-2} \om_k = - \langle\rho\rangle^{(0)}. \ee To exhibit more clearly the dependence  of this adiabatic order on $C$,  by
rescaling the integration variable with a factor $C^{-1/2}$, we rewrite it as \be
\langle\rho\rangle^{(0)}=-\langle p\rangle^{(0)}
=\frac{\Omega_{n-1}\mu^{4-n}}{2\,(2\pi)^{n-1}}\int dk\,k^{n-2}
\tilde{\om}_k  ,\label{ZeO} \ee where
$\tilde{\om}_k=\om_k/\sqrt{C}$.
Thus, as we are working with the metric in the conformal form, we
see that $\langle T_{\mu\nu}\rangle^{(0)} =  N_0 g_{\mu\nu}$ (with
$N_0$ a divergent factor) so that  we can define $\langle
\widetilde{T}_{\mu\nu}\rangle=\langle T_{\mu\nu}\rangle-\langle
T_{\mu\nu}\rangle^{(0)}$ and the SEE
\begin{equation}
G_{\mu\nu}+\Lambda_{\mathrm
B}g_{\mu\nu}=8\pi G(\langle T_{\mu\nu}\rangle - \langle T_{\mu\nu}\rangle^{(0)} + \langle
T_{\mu\nu}\rangle^{(0)} )
\end{equation}
can be recast in the form
\begin{equation}
G_{\mu\nu}+\Lambda_{\mathrm R}g_{\mu\nu}=8\pi G\langle
\widetilde{T}_{\mu\nu}\rangle
\end{equation}
where $\Lambda_{\mathrm R}= \Lambda_{\mathrm B}-8\pi G N_0$ is the
renormalized cosmological constant.

Since in the case of a generalized dispersion relation for which
$\om_k^2\sim k^r$ with $r\geq 8$, the energy momentum tensor can
be renormalized by substracting the zeroth adiabatic order, we can
make the identification $\langle T_{\mu\nu}\rangle_{\mathrm{
Ren}}\equiv \langle \widetilde{T}_{\mu\nu}\rangle$ (i.e., $\langle
\rho \rangle_{\mathrm{Ren}}= \langle \rho \rangle - \langle \rho
\rangle^{(0)}$ and $\langle p \rangle_{\mathrm{Ren}} = \langle p \rangle -
\langle p \rangle^{(0)}$). In such a case, as these expressions
are already finite, in order to evaluate them in terms of the
modes of the scalar field it is not necessary to work in $n$
dimensions: one can first take the limit  $n\to 4$ and then
perform the momentum integration, that is
\begin{eqnarray}
\nonumber\langle\rho\rangle_{\mathrm{Ren}} &=& \frac{1}{ C^2}\int
\frac{d^3k}{(2\pi)^{3}} \left\{ \frac{C}{2}\left|\lp
\frac{\chi_k}{\sqrt{C}}\rp' \right|^2
+\frac{\om_k^2}{2}\,|\chi_k|^2+\xi G_{\eta\eta}|\chi_k|^2\right.
\\ &+&\left.\frac{3}{2}\xi
\lb\frac{C'}{C}(\chi_k'\chi_k^*+\chi_k{\chi_k'}^{*})-\frac{{C'}^2}{C^2}
|\chi_k|^2 \rb -\frac {1}{4} \om_k \right\}\\ \nonumber\langle
p\rangle_{\mathrm{Ren}} &=& \frac{1}{C^2}\int \frac{d^3k}{(2\pi)^{3}}
\left\{\lp\frac{1}{2}-2\xi\rp C\left|\lp
\frac{\chi_k}{\sqrt{C}}\rp' \right|^2+\xi
G_{11}|\chi_k|^2\right.\\\nonumber
 &+&\lb \lp\frac{k^2}{3}\rp\frac{d\om^2}{
dk^2}-\frac{\om_k^2}{2}\rb|\chi_k|^2-\xi(\chi_k''\chi_k^*+\chi_k{\chi_k''}^{*})+\xi\frac{
C'}{2C}(\chi_k'\chi_k^*+\chi_k{\chi_k'}^{*})\\
&-&\left.\xi\lp\frac{C''}{C}-\frac{{C'}^2}{C^2}\rp|\chi_k|^2 -
\frac{k^2}{6\om_k}\frac{d\om^2_k}{ dk^2} \right\}.
\end{eqnarray}
The adiabatic renormalization procedure works only for the vacuum
states of the field that coincide with the adiabatic vacuum up to
the order subtracted \cite{Cast}. If we assume that the scalar
field is in  the vacuum state near the initial singularity $C\to
0$, this means that the exact solution $\chi_k$ of Eq.(\ref{PXXP})
should coincide with the WKB solution up to that order for $C\to
0$. This fact ensures that the above integrals are finite.

\subsection{Second adiabatic order}
So far we have shown that the zeroth adiabatic order of the vacuum
expectation values of the energy density and pressure satisfy Eq.
(\ref{ZeO}) and, hence, that they can be absorbed by a
redefinition of the cosmological constant. In what follows, we
shall see that the WKB expansion of $\langle\rho\rangle$ and
$\langle p\rangle$, up to second adiabatic order, have the
appropriate structure required to remove the divergences in the
stress tensor that appear in the cases $\om_k^2\sim k^6$ and
$\om_k^2\sim k^4$. Therefore, all divergences will be absorbed
renormalizing  the cosmological and Newton constants in the SEE.
More specifically, with the use of dimensional regularization, we
shall show that the second adiabatic orders of
$\langle\rho\rangle$ and $\langle p \rangle$ are proportional to
the components $G_{\eta\eta}$ and $G_{1 1}$ of the Einstein
tensor, respectively, yielding a renormalization of the Newton
constant.

Using integration by parts and some algebra one can find
expressions for $\langle \rho\rangle^{(2)}$ and $\langle
p\rangle^{(2)}$ that involve only the two integrals
\begin{eqnarray}
I_1&=& \int dx\,\frac{x^{\frac{n-3}{2}}}{\tilde{\om}_k}\nonumber \\
I_2&=& \int
dx\,\frac{x^{\frac{n+1}{2}}}{\tilde{\om}_k^3}\frac{d^2\tilde{\om}_k^2}{{d
x}^2}\label{integrals}
\end{eqnarray}where $x\equiv k^2/C$ and, as above, $\tilde{\om}_k=\om_k/\sqrt{C}$.

As an example, let us consider the following integral
\begin{eqnarray}
 I\equiv\int d k\, \frac{k^{n-2}}{C^{\frac{n-2}{2}}\om_k}\lp 1-\frac{k^2}{\om_k^2}\frac{d{\om}_k^2}{d k^2} \rp^2,
  \end{eqnarray}
 which contributes to both $\langle\rho\rangle^{(2)}$ and  $\langle p\rangle^{(2)}$.
In order to rewrite it we can proceed as follows
\begin{eqnarray}
 I&=&\frac{1}{2}\int d x\,
\frac{ x^{\frac{n-3}{2}}}{\tilde{\om}_k}+2\int d x\,
x^{\frac{n-1}{2}}\frac{d\tilde{\om}_k^{-1}}{d x\,\,\,\,\,\,\,
}-\frac{1}{3}\int d x x^{\frac{n+1}{2}}
\frac{d\tilde{\om}_k^{-3}}{d x\,
\,\,\,\,\,\,}\frac{d\tilde{\om}_k^{2}}{d x\,\,\,\,}
\nonumber\\
&=&\frac{[(n-3)^2-1]}{6}\int d x\,
\frac{x^\frac{n-3}{2}}{\tilde{\om}_k}+\frac{1}{3}\int d x\,
\frac{x^{\frac{n+1}{2}}}{\tilde{\om}_k^3}\frac{d^2\tilde{\om}_k^2}{{dx\,}^2\,\,}=\frac{[(n-3)^2-1]}{6}\,I_1+\frac{1}{3}
I_2,
\end{eqnarray}
where the first equality follows after the change of variables
$x=k^2/C$ and some rearrangements of the integrand, while the
second one is obtained, with the use of dimensional
regularization, after two integrations by parts.

Applying a similar procedure to the other integrals we get, after
a long calculation,
\begin{eqnarray}
\langle\rho\rangle^{(2)}_{\xi=0} & = &\frac{G_{\eta\eta}}{C}\frac{\Omega_{n-1}\,\mu^{4-n}}{4\,(2\pi)^{n-1}}\left\{-\frac{[n+2+n (n-4)]}{6(n-1)(n-2)}I_1\right.\nonumber\\
&+&\left.\frac{1}{6(n-1)(n-2)}I_2\right\},\label{rhoxi}\\
\langle p\rangle^{(2)}_{\xi=0}
& =& \langle\rho\rangle^{(2)}_{\xi=0}\lb 1-2\frac{(n-4)}{(n-1)}-\frac{4}{(n-1)}\frac{C''}{C}\frac{C^2}{{C'}^2}\rb= \frac{G_{1 1}}{G_{\eta\eta}} \langle\rho\rangle^{(2)}_{\xi=0}\\
\langle\rho\rangle^{(2)}_{\xi} & = & \frac{\xi G_{\eta\eta}}{
C}\frac{\Omega_{n-1}\,
\mu^{4-n}}{4\, (2\pi)^{n-1}} I_1 ,\label{rhoxixi}\\
\langle p\rangle^{(2)}_{\xi} & =& \frac{\xi G_{1 1}}{
C}\frac{\Omega_{n-1}\,\mu^{4-n}}{4\,(2\pi)^{n-1}}I_1,\label{pxi}
\end{eqnarray}
Notice that in the case of the usual dispersion relation, the
integral containing the second derivative of $\om_k^2$ vanishes
and we recover the known second adiabatic order results
\cite{bunch}.
It is worth mentioning here that, for $\om_k^2\sim k^6$, the
leading terms in the second adiabatic order cancel out for
large $k$, and thus for minimal coupling $\xi=0$ the second
adiabatic order is finite as $n \to 4$. This can be seen directly
from Eq. (\ref{rhoxi}) or from Eq. (\ref{rhominimcoup}).

Eqs. (\ref{rhoxi})-(\ref{pxi}) above are enough for our purposes,
since it is clear from them that the second adiabatic order of
$\langle\rho\rangle$ and $\langle p\rangle$ are proportional to
$G_{\eta\eta}$ and $G_{11}$ respectively.
If we write $\langle T_{\mu\nu}\rangle^{(n)}$ for the term of
adiabatic order $n$ of the stress-tensor, we find that \be \langle
T_{\mu\nu}\rangle^{(0)} =  N_0 g_{\mu\nu} \ee where  $N_0$ is in
principle a divergent factor, so that the corresponding
contributions can be removed by introducing a renormalized
cosmological constant $\Lambda_{\mathrm R}$. On the other hand, we
have \be \langle T_{\mu\nu}\rangle^{(2)}  =  N_2 G_{\mu\nu}, \ee
where $N_2$ is another divergent factor, and hence these
contributions can be absorbed in a renormalization of the Newton
gravitational constant. Therefore we can define \be \langle
T_{\mu\nu}\rangle_{\mathrm{Ren}}=\langle T_{\mu\nu}\rangle-
\langle T_{\mu\nu}\rangle^{(0)}-\langle T_{\mu\nu}\rangle^{(2)}
\ee and write the SEE as \be G_{\mu\nu}+\Lambda_{\mathrm
R}g_{\mu\nu}=8\pi G_{\mathrm R}\langle
T_{\mu\nu}\rangle_{\mathrm{Ren}}. \ee Differing from the case of
standard dispersion relations $\om_k^2\sim k^2$, now  all
contributions of adiabatic orders higher than the second are
finite, so that  no additional terms must be included in the SEE
in order to deal with physically meaningful quantities.

\subsection{Evaluation of the regularized integrals}

The explicit expression for the constants $N_0$ and $N_2$ can be
obtained  by a direct computation in Eqs.
(\ref{rhoderiv})-(\ref{pderiv}) or equivalently from Eqs.
(\ref{ZeO}) and (\ref{rhoxi})-(\ref{pxi}).

As a first example, let us consider the case of a massless field
with a dispersion relation of the form
$\omega_k^2=k^2+2b_{11}k^4/C+2|b_{12}|k^6/C^2$ (we will assume
that $|b_{12}|>b_{11}^2/2$ to avoid zeros of the frequency). In
this case the divergent contributions come from the zeroth
adiabatic order and from the terms proportional to $\xi$  of the
second one, since (as we already mentioned) the second adiabatic
order is finite for $\xi=0$. Then, after computing the integrals
we obtain \cite{GR}
\begin{eqnarray}&&
 \langle \rho \rangle^{(0)}  =  - \langle p \rangle^{(0)}
=\frac{\mu^{4-n}(2
|b_{12}|)^{\frac{(2-n)}{4}}}{4\Gamma\lb-\frac{1}{2}\rb(4\pi)^{\frac{(n-1)}{2}}}
 \left\{\frac{1}{\sqrt{2 |b_{12}|}}\, \Gamma \lb
\frac{n}{4}\rb \Gamma\lb -\frac{1}{2}-\frac{n}{4} \rb
\right. \,\,\,\,\,\,\,\,\,\,\,\,\,\,\,\,\,\,\,\,\,\,\,\,\,\,\,\,\,\,\,\nonumber\\
&&\left.\times \,\, _2F_1\lb
-\frac{1}{2}-\frac{n}{4},\frac{n}{4};\frac{1}{2};\frac{b_{11}^2}{2
|b_{12}|}\rb-\frac{b_{11}}{|b_{12}|}\,
\Gamma\lb\frac{n}{4}+\frac{1}{2}\rb \Gamma\lb-\frac{n}{4}\rb\,
_2F_1\lb\frac{n}{4}+\frac{1}{2},-\frac{n}{4};\frac{3}{2};\frac{b_{11}^2}{2|b_{12}|}\rb
\right\},\\&&
 \langle \rho \rangle^{(2)}_{\xi=0}  = \langle p \rangle^{(2)}_{\xi=0} \frac{G_{\eta\eta}}{G_{11}}=
 \frac{G_{\eta\eta}}{C}
\frac{\mu^{4-n} (2
|b_{12}|)^{\frac{(2-n)}{4}}}{24(4\pi)^{\frac{(n-1)}{2}}}
 \left\{\frac{b_{11}}{\sqrt{2 |b_{12}|}}\,
\frac{\Gamma \lb \frac{1}{2}-\frac{n}{4}\rb \Gamma\lb\frac{n}{4}
\rb}{4\Gamma\lb\frac{1}{2}\rb\Gamma\lb\frac{n}{2}+\frac{1}{2}\rb}
\right. \,\,\,\,\,\,\,\,\,\,\,\,\,\,\,\,\,\,\,\,\,\,\,\,\,\,\,\,\,\,\,\,\,\,\,\,\,\,\,\,\,\,\,\,\,\,\,\,\,\,\,\,\,\,\,\,\,\,\,\,\,\,\,\,\,\,\,\,\,\,\,\,\,\,\,\,\,\,\,\,\,\,\,\nonumber\\
&&\times \lp _2F_1\lb
\frac{3}{2}-\frac{n}{4},\frac{n}{4};\frac{1}{2};\frac{b_{11}^2}{2
|b_{12}|}\rb-\frac{(n-4)(n-2)}{2}\,
_2F_1\lb\frac{3}{2}-\frac{n}{4},\frac{n}{4};\frac{3}{2};\frac{b_{11}^2}{2|b_{12}|}\rb\rp\nonumber\\
&&\left.+ \frac{\Gamma \lb 2-\frac{n}{4}\rb
\Gamma\lb\frac{n}{4}-\frac{1}{2}\rb}{\Gamma\lb\frac{1}{2}\rb\Gamma\lb\frac{n}{2}+\frac{1}{2}\rb}\lp
 _2F_1\lb
1-\frac{n}{4},\frac{n}{4}-\frac{1}{2};\frac{1}{2};\frac{b_{11}^2}{2
|b_{12}|}\rb+\frac{b_{11}^2}{4|b_{12}|}\, _2F_1\lb
2-\frac{n}{4},\frac{n}{4}+\frac{1}{2};\frac{3}{2};\frac{b_{11}^2}{2|b_{12}|}\rb\rp\right\},\end{eqnarray}
\begin{eqnarray} &&
 \langle \rho \rangle^{(2)}_{\xi}  = \langle p \rangle^{(2)}_{\xi} \frac{G_{\eta\eta}}{G_{11}}= \xi
 \frac{G_{\eta\eta}}{C}
\frac{\mu^{4-n} (2
|b_{12}|)^{\frac{(4-n)}{4}}}{4\Gamma\lb\frac{1}{2}\rb(4\pi)^{\frac{(n-1)}{2}}}
 \left\{\frac{1}{\sqrt{2 |b_{12}|}}\,
\Gamma \lb 1-\frac{n}{4}\rb \Gamma\lb -\frac{1}{2}+\frac{n}{4} \rb
\right. \,\,\,\,\,\,\,\,\,\,\,\,\,\,\,\,\,\,\,\,\,\,\,\,\,\,\,\,\,\,\,\,\,\,\,\,\,\,\,\,\,\,\,\,\,\,\,\,\,\,\,\,\,\,\,\,\,\,\,\,\,\,\,\,\,\,\,\,\,\,\nonumber\\
&&\left.\times \,\, _2F_1\lb
1-\frac{n}{4},-\frac{1}{2}+\frac{n}{4};\frac{1}{2};\frac{b_{11}^2}{2
|b_{12}|}\rb-\frac{ b_{11}}{|b_{12}|}\,
\Gamma\lb\frac{3}{2}-\frac{n}{4}\rb \Gamma\lb\frac{n}{4}\rb\,
_2F_1\lb\frac{3}{2}-\frac{n}{4},\frac{n}{4};\frac{3}{2};\frac{b_{11}^2}{2|b_{12}|}\rb
\right\}
 \end{eqnarray} where we have used some properties of
the gamma $\Gamma$ and hypergeometric $_2F_1$ functions
\cite{Abram}.

The behavior of $\langle \rho \rangle^{(0)}$ and
$\langle \rho \rangle^{(2)}$ in the limit $n \to 4$ is
\begin{eqnarray}
 \langle \rho
\rangle^{(0)}&=&-\frac{b_{11}(b_{11}^2-2|b_{12}|)}{32\sqrt{2}\pi^2
|b_{12}|^{5/2}}
\lb\frac{1}{n-4}-\ln(|b_{12}|^{1/4}\mu)\rb+\mathcal{O}(n-4),\\
 \langle \rho\rangle^{(2)}&=&-\frac{\xi G_{\eta\eta}}{C (2\pi)^2 \sqrt{2 |b_{12}|}}\lb\frac{1}{n-4}-\ln({|b_{12}|^{1/4}\mu})\rb+\lp\hbox{finite $\mu$-independent terms  as }n \to 4\rp,
\end{eqnarray}  where we have redefined $\mu$ to
absorb a constant term.

Let us now consider a dispersion relation of the form
$\om_k^2=k^2+C m^2+2 b_{11} k^4/C$, with $b_{11}>0$. The integrals in Eqs.
(\ref{ZeO}) and (\ref{rhoxi})-(\ref{pxi}) can be computed
explicitly. Recalling some properties of the Gamma and
hypergeometric functions \cite{Abram}, in the limit $n\to 4$ we
obtain
\begin{eqnarray}
\langle\rho\rangle^{(0)} & = &-\langle p\rangle^{(0)}=\frac{m^{3/2}}{ 2^{1/4}64 b_{11}^{5/4}\pi^{5/2}}\left\{\Gamma\lb-\frac{3}{4}\rb\Gamma\lb\frac{5}{4}\rb\, _2F_1\lb-\frac{3}{4},\frac{5}{4};\frac{3}{2};\frac{1}{8 b_{11} m^2}\rb\right.\nonumber\\
&-&\left.\sqrt{2 b_{11} m^2}\,\Gamma\lb-\frac{5}{4}\rb\Gamma\lb\frac{3}{4}\rb\, _2F_1\lb-\frac{5}{4},\frac{3}{4};\frac{1}{2};\frac{1}{8 b_{11} m^2}\rb\right\},\\
\langle\rho\rangle^{(2)}_{\xi=0} & = & \langle p \rangle^{(2)}_{\xi=0} \frac{G_{\eta\eta}}{G_{11}}=\frac{\sqrt{m}}{2^{3/4} 2304 b_{11}^{3/4}\pi^{5/2}}\left\{4\sqrt{8 b_{11} m^2}\, \Gamma\lb\frac{1}{4}\rb^2\lb -2\frac{(1-12 b_{11} m^2)}{1-8 b_{11} m^2}\,\right.\right.\nonumber\\
&\times& \left._2F_1\lb-\frac{3}{4},\frac{1}{4};\frac{1}{2};\frac{1}{8 b_{11} m^2}\rb+3\, _2F_1\lb\frac{1}{4},\frac{1}{4};\frac{1}{2};\frac{1}{8 b_{11} m^2}\rb\rb\nonumber\\
&+&\left.\Gamma\lb-\frac{1}{4}\rb^2\lb 2\, _2F_1\lb-\frac{1}{4},\frac{3}{4};\frac{1}{2};\frac{1}{8 b_{11} m^2}\rb+\,_2F_1\lb\frac{3}{4},\frac{3}{4};\frac{1}{2};\frac{1}{8 b_{11} m^2}\rb\rb\right\},\\
 \langle \rho \rangle^{(2)}_{\xi}& = &\langle p \rangle^{(2)}_{\xi} \frac{G_{\eta\eta}}{G_{11}}=
\frac{1}{2^{1/4} 32 b_{11}^{5/4}\sqrt{m}\pi^{5/2}}\left\{\sqrt{2 b_{11} m^2}\, \Gamma\lb-\frac{1}{4}\rb\Gamma\lb\frac{3}{4}\rb\,\right. \nonumber\\ &\times &\left. _2F_1\lb-\frac{1}{4},\frac{3}{4};\frac{1}{2};\frac{1}{8 b_{11} m^2}\rb
-\Gamma\lb\frac{1}{4}\rb\Gamma\lb\frac{5}{4}\rb\,
_2F_1\lb\frac{1}{4},\frac{5}{4};\frac{3}{2};\frac{1}{8 b_{11}
m^2}\rb\right\}.
\end{eqnarray}
In the case of a massless field, the results simplify to
\begin{eqnarray}
\langle\rho\rangle^{(0)} & = &-\langle p\rangle^{(0)} = \frac{1}{ 32\pi^2 b_{11}^2}\frac{\Gamma (-5/2)}{ \Gamma (-1/2)}=\frac{1}{120 b_{11}^2 \pi^2},\\
\langle\rho\rangle^{(2)} & = -& \ \frac{(1-18\xi)}{ 288\pi^2 b_{11}}\frac{\Gamma (-1/2)}{ \Gamma (1/2)}\frac{G_{\eta\eta}}{C}= \frac{(1-18\xi)}{144\pi^2 b_{11}}\frac{G_{\eta\eta}}{C},\\
\langle p\rangle^{(2)} & = & -\frac{(1-18\xi)}{288\pi^2 b_{11}}
\frac{\Gamma (-1/2)}{ \Gamma (1/2)}\frac{G_{11}}{C}=
\frac{(1-18\xi)}{144\pi^2 b_{11}}\frac{G_{11}}{C}.
\end{eqnarray}
All the contributions which are in principle divergent give finite results when evaluated by means of dimensional regularization (all negative arguments appearing in the gamma functions are non-integer). This could have been anticipated, because the dependence $\om_k^2\sim k^4$ of the dispersion relation leads to integral expressions which are formally equivalent (for large $k$ or in the massless limit) to which would be obtained in $2+1$ dimensions for a standard dispersion relation, and  in this case dimensional regularization leads to finite results for integrals which are in principle divergent \cite{3d}.

\section{Conclusions}

 We have given a prescription for obtaining finite, physically meaningful, expressions
 for the components of the stress tensor for a field of arbitrary coupling,
 with generalized dispersion relations, in a FRW background. We have followed the
 usual procedure of  subtracting  from the exact components of the  energy-momentum
 tensor the, in principle, divergent contributions of the corresponding  expressions
 obtained from the adiabatic expansion, which we have  evaluated by means of
 dimensional regularization.

 We have seen that, differing from the usual case
 corresponding to standard dispersion relations with $\om^2_k\sim k^2$, the
 fourth adiabatic order is convergent. Consequently,
 additional terms proportional to the geometric tensors $H_{\mu\nu}^{(1)}$ and
 $H_{\mu\nu}^{(2)}$ associated with  corrections of second order in the curvature
 are not necessary in the SEE, and  the renormalization  does not require more
 than  the redefinition of  the cosmological constant and the Newton
constant. At first sight it may look surprising that for
generalized  dispersion relations with $\om^2_k\sim k^r, r\geq 4$
the divergences are milder than for the standard case. The reason
is that, while the divergence of the zeroth adiabatic order is
stronger, the higher orders are suppressed by powers of
$\om_k^{-2}$. Therefore, for the cases $\om_k^2\sim k^6$ and
$\om^2_k\sim k^4$, the fourth adiabatic orders are already finite.

 In the case of dispersion relations of the form $\om^2_k\sim k^r$ with $r\geq 8$ or
 $\om^2_k\sim k^6$ with
 $\xi=0$, the second adiabatic order is finite and the divergences are contained
 in the zeroth adiabatic
 order.
 Therefore, in such cases, the adiabatic renormalization is equivalent to the
 subtraction of the zero point energy of each field Fourier mode,
 as done in Ref. \cite{branma}.

There are several issues that would deserve further investigation.
From a formal point of view, it would  be interesting to extend
the renormalization of the stress tensor to interacting theories
with non standard dispersion relations. From a "phenomenological"
point of view, the renormalized SEE obtained in this paper should
be the starting point to evaluate whether the backreaction of
trans-Planckian modes prevents inflation or not.

\section{Appendix: Renormalization of $\langle\phi^2\rangle$}

 We shall consider the simple problem of the  mean squared field in order to illustrate
 how one can effectively make the subtraction leading to  a finite quantity starting from
 a divergent integral. We will consider the dispersion relation  (\ref{dis}) in the
 particular case
 $\om_k^2\sim k^4$  for a De Sitter  evolution  $C(\eta)=\alpha^2/\eta^2$.
The equation for the  associated field modes $\ck$ reads \be
\frac{\partial ^2\ck}{ \partial \eta^2}+\lp k^2+\frac{\tilde\mu^2
\alpha^2}{\eta^2}+\frac{2b_{11}k^4 \eta^2}{\alpha^2}\rp\ck=0, \ee
where $\tilde\mu^2=m^2+n(n-1)(\xi-\xi_n)/\alpha^2$. This equation
can be solved exactly in the case $\tilde\mu^2=0$ or for arbitrary
values of $\tilde\mu^2$ in the limit $\eta\to - \infty$. Indeed,
in both cases, with the substitution
$s=(2b_{11})^{1/4}\alpha^{-1/2}k\eta$ and introducing the constant
$\la=\alpha (2b_{11})^{-1/2}$, the equation to be solved becomes
\be \frac{\partial ^2\ck}{ \partial s^2}+\lp \la+ s^2\rp\ck=0. \ee
The solution is of the parabolic form \be \ck (s)=
D_{-\lp\frac{1+i\la}{ 2}\rp}\lb\pm(1+i)s\rb \ee (see Ref.
\cite{GR} for the definition of the parabolic function $D$).

For our purposes it is enough with the expansion for large $|s|$,
which corresponds to $\eta\to - \infty$; this expansion has the
form \be D_p(z)\approx e^{-z^2/4}z^p\lp 1-\frac{p(p-1)}{
2z^2}+\frac{p(p-1)(p-2)(p-3)}{ 8 z^4}- \cdots \rp,
\label{expmodos}\ee where $p=-(1+i\la)/2$ and $z=\pm (1+i)s$.
After imposing the normalization condition (\ref{nor}), by power
counting (and recalling that $s\sim \eta k$)  it is easy to see
that the only divergence comes from the leading order term of the
expansion. This term is \be \ck^{(0)}=\frac{1}{
\sqrt{k}}\lp\frac{\la}{
2}\rp^{1/4}\lp\sqrt{2}s\rp^{-\lp\frac{1+i\la}{ 2}\rp} \exp
\lb-\frac{i}{2} \lp\frac{\pi}{ 4}+s^2\rp\rb, \ee so that \be
\left|\ck^{(0)}\right|^2=\frac{\alpha}{2k^2|\eta|\sqrt{2 b_{11}}}
\ee
 and when substituted in the integral for the vacuum expectation value
\be \langle\phi^2\rangle =
\frac{\sqrt{C}}{2}\frac{\Omega_{n-1}\mu^{4-n}}{(2\pi\sqrt{C})^{n-1}}\int
dk\,k^{n-2} \left|\ck\right|^2\label{phidos} \ee gives a linear
divergence. The contribution of the other terms of the expansion
Eq. (\ref{expmodos}) is finite.

For the WKB solutions we have $|\ck |^2 =\ck\ck^*  =  1/ (2 W_k)$.
Because for large values of $k$ we have $\om_k^2\sim k^4$, then
the only divergence appearing in the inner product comes from the
lowest order of the expansion. This contribution is given by \be
\langle\phi^2\rangle^{(0)} =
\frac{\sqrt{C}}{2}\frac{\Omega_{n-1}\mu^{4-n}}{(2\pi\sqrt{C})^{n-1}}\int
\frac{dk\,k^{n-2}} {2\om_k}\label{phi0} \ee which clearly diverges
linearly. Once this   quantity is calculated,  then the
renormalized mean squared field can be defined by subtracting the
adiabatic expansion from the exact result: \be
\langle\phi^2\rangle_{\mathrm{Ren}}=\langle\phi^2\rangle
-\langle\phi^2\rangle^{(0)} \ee For the sake of illustration, let
us evaluate explicitly Eq.(\ref{phi0}) in $n$ dimensions \be
\langle\phi^2\rangle^{(0)} =
\frac{\sqrt{C}}{4}\frac{\Omega_{n-1}\mu^{4-n}}{(2\pi\sqrt{C})^{n-1}}\int
dk\,k^{n-3}\lp \frac {C}{2b_{11}}\rp^{1/2}\frac{1}{\lp
k^2+\frac{C}{2 b_{11}}\rp^{1/2}}\label{phi2}. \ee Note that  (up
to the factor $(C/2b_{11})^{1/2}$) Eq. (\ref{phi2}) is formally
analogous, in the limit $n\to 4$, to what one would obtain in the
case  of a field of nonvanishing ``mass'' given by $m^2=1/2b_{11}$
in 2+1 dimensions. Thus we  expect a finite result after  applying
the usual formulae for dimensional regularization \cite{3d}.
Indeed, we obtain \be \langle\phi^2\rangle^{(0)} = \frac{1}{16
b_{11}}\frac{\Gamma(-1/2)}{\Gamma(1/2)} = -\frac{1}{8 b_{11}}, \ee
which is finite.

To obtain the finite result for
$\langle\phi^2\rangle_{\mathrm{Ren}}$ one should be able to
compute the integral of the modes in Eq. (\ref{phidos}), which
should also be finite in dimensional regularization. This is not
appealing from a practical point of view, since in general there
will be no analytical expression for this integral. However, as
the difference \be \langle\phi^2\rangle_{\mathrm{Ren}}
=\frac{\sqrt{C}}{2}\frac{\Omega_{n-1}\mu^{4-n}}{(2\pi\sqrt{C})^{n-1}}\int
dk\,k^{n-2}\left[ \left|\ck\right|^2-\frac{1}{2\om_k}\right] \ee
is convergent , one can take the limit $n\to 4$ inside the
integral and evaluate numerically both $\ck$ and the momentum
integral already at $n=4$.

\begin{acknowledgments}

This work has been supported by  Universidad de Buenos Aires,
CONICET and ANPCyT.

\end{acknowledgments}


\begin{thebibliography}{99}
\bibitem{foot}To renormalize $\langle\phi^2\rangle$ it is necessary
to consider the Hadamard two point function truncated at the
second adiabatic order.
\end{thebibliography}
\end{document}